\documentclass[graybox]{svmult}

\usepackage{type1cm}        
\usepackage{makeidx}         
\usepackage{graphicx}        
\usepackage{multicol}        
\usepackage[bottom]{footmisc}

\usepackage{newtxtext}       %
\usepackage{newtxmath}       

\makeindex             

\usepackage{orcidlink}
\usepackage{todonotes}
\graphicspath{{figs/}}
\usepackage{kbordermatrix}
\usepackage{blkarray}
\usepackage{tabularray}
\UseTblrLibrary{booktabs}
\usepackage{verbatim}
\usepackage{hyperref}

\hypersetup{
	pdftitle={Bayesian Estimation of Laser Linewidth from Delayed Self-Heterodyne Measurements},
	pdfauthor={Lutz Mertenskötter and Markus Kantner},
	pdfkeywords={laser linewidth, phase noise, narrow-linewidth semiconductor lasers, self-heterodyne beat note measurement, non-Markovian noise, Langevin equations, stochastic differential equations, Bayesian inference, Markov chain Monte Carlo, Metropolis-Hastings, time series analysis, spectral analysis},
	pdfsubject = {},
	breaklinks=true,
	colorlinks=true,
	citecolor=blue,
	linkcolor=blue,
	urlcolor=blue,
}

\begin{document}

\title*{Bayesian Estimation of Laser Linewidth from Delayed Self-Heterodyne Measurements}
\author{Lutz Mertenskötter \orcidlink{0000-0001-5074-2721} and Markus Kantner \orcidlink{0000-0003-4576-3135}}
\authorrunning{L. Mertenskötter and M. Kantner}
\institute{L.~Mertenskötter~$\cdot$~M.~Kantner
\at Weierstrass Institute for Applied Analysis and Stochastics, Mohrenstr. 39, 10117 Berlin, Germany,
email: \href{mertenskoetter@wias-berlin.de}{mertenskoetter@wias-berlin.de}, \href{kantner@wias-berlin.de}{kantner@wias-berlin.de}
}
%
%
\maketitle

\abstract{We present 
a statistical inference approach to estimate
the frequency noise characteristics of ultra-narrow linewidth lasers from delayed self-heterodyne beat note measurements using Bayesian inference. Particular emphasis is on estimation of the intrinsic (Lorentzian) laser linewidth.
The approach is based on a statistical model of the measurement process, taking into account the effects of the interferometer as well as the detector noise.
Our method therefore yields accurate results even when the intrinsic linewidth plateau is obscured by detector noise.
The regression is performed on periodogram data in the frequency domain using a Markov-chain Monte Carlo method.
By using explicit knowledge about the statistical distribution of the observed data, the method yields  good results already from a single time series and does not rely on averaging over many realizations, since the information in the available data is evaluated very thoroughly.
The approach is demonstrated for simulated time series data from a stochastic laser rate equation model with $1/f$--type non-Markovian noise.
}

\section{Introduction}
\label{sec:Introduction}
Ultra-narrow linewidth lasers are critical components of many applications
in modern science and technology, ranging from precision metrology,
\emph{e.g.}, gravitational wave interferometers \cite{Abbott2009} and
optical atomic clocks \cite{Ludlow2015}, to coherent optical communication 
systems \cite{Kikuchi2016} and ion-trap quantum computers \cite{Akerman2015}.
To perform well in these technologies, the laser requires a high degree of
spectral coherence, \emph{i.e.}, a well-defined phase, and/\,or a sharply
defined frequency and thus low frequency noise. The (intrinsic) laser
linewidth is quantified by the width of the optical power spectrum, which
in real systems is usually broadened by additional $1/f$--like technical noise
(also \emph{flicker noise}) \cite{Kikuchi1989,Mercer1991,Stephan2005}.
As the width of the optical power spectrum is dominantly determined by the
frequency noise, the corresponding frequency noise power spectral
density (FN--PSD) provides an almost complete characterization of the
spectral quality of the laser, where the effects of non-Markovian noise
can be well separated from that of white noise. The latter manifests itself
as a plateau in the high-frequency part of the FN--PSD, from which the
intrinsic (Lorentzian) laser linewidth \cite{Henry1986, Wenzel2021} can be
deduced, that is of major interest for most of the aforementioned applications.
In ultra-narrow linewidth lasers, the determination of the white noise
plateau can be challenging, since it often sets in only at very high frequencies
and is obscured by $1/f$--type noise (at low frequencies) or by detector noise
(at high frequencies).

The standard technique for the experimental measurement of the laser linewidth
is the delayed self-heterodyne (DSH) method \cite{Okoshi1980, Schiemangk2014},
which involves measuring the beat signal between the optical field with a delayed
and frequency-shifted copy of itself. This method is attractive because it can
provide a direct measurement of the linewidth without the need for an external
frequency standard or active frequency stabilization. Evaluation of the DSH
measurement data is however non-trivial, as both the footprint of the interferometer
as well as the detector noise must be removed in order to obtain an artifact-free reconstruction
of the FN--PSD \cite{Kantner2023}.

In this paper, we present a Bayesian estimation approach to infer on the
laser's FN--PSD from time series data. Our method is based on statistical
modeling of the DSH measurement process and allows to extract accurate estimates
of key parameters such as the intrinsic linewidth, even when the white noise
plateau is obscured by detector noise. The method is demonstrated for simulated
time series data based on a stochastic rate equation model for a single-mode
semiconductor laser.

\section{Delayed Self-Heterodyne Method}
\label{sec:DSH}

\subsection{Experimental Setup}

\begin{figure}[t]
\centering
\includegraphics[scale=1.0]{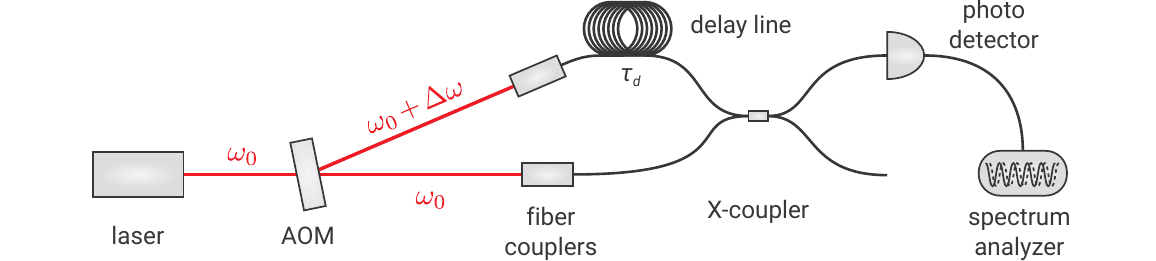}
\caption{Experimental setup of the DSH method, where an acousto-optic
modulator (AOM) separates the incoming laser beam into two beams. One part of the
signal is delayed by a long fiber and frequency shifted. The two beams are
superimposed at a photodetector, which captures only the slow beat note signal.
The picture is reprinted with permission from Ref.~\cite{Kantner2023}.}
\label{fig:DSHsetup}
\end{figure}

The DSH method is a standard technique for measuring the laser
linewidth \cite{Schiemangk2014}, particularly of ultra-narrow linewidth lasers.
It involves splitting of the laser beam into two paths using an
acousto-optic modulator (AOM), where one of the two beams is also frequency-shifted,
see Fig.~\ref{fig:DSHsetup}. The frequency-shifted beam is then delayed using
a long fiber before being finally superimposed with the other beam at a photodetector.
The optical field received by the detector
$E\left(t\right)=\mathrm{Re}\left(\mathcal{E}\left(t\right)\right)$ reads
\begin{equation}
\label{eq:measuredE}
\mathcal{E}\left(t\right)=\sqrt{P\left(t\right)}\,\mathrm{e}^{i\left(\omega_{0}t+\phi\left(t\right)\right)}+\sqrt{P\left(t-\tau_{d}\right)}\,\mathrm{e}^{i\left(\left(\omega_{0}+\Delta\omega\right)\left(t-\tau_{d}\right)+\phi\left(t-\tau_{d}\right)\right)}+\xi_{E}\left(t\right),
\end{equation}
where $\omega_0$ is the nominal continuous wave (CW) frequency, 
$\tau_d$ is the interferometer delay, $\phi\left(t\right)$ is the (randomly
fluctuating) optical phase, $P\left(t\right)$ is the (fluctuating) amplitude
and $\Delta\omega$ is the frequency shift induced by the AOM.
Moreover, $\xi_{E}\left(t\right)$ describes additive detector noise.
Conventional photodetectors capture only the slow beat note in the
intensity signal $I\left(t\right)\propto\left|E\left(t\right)\right|^{2}$.
The spectrum analyzer then further downshifts the beat frequency and removes
the DC component of the signal. From this signal and its quadrature component
(generated by a Hilbert transform), one can then extract the phase jitter and
finally conclude on the fluctuations of the instantaneous laser frequency \cite{Kantner2023}.

\subsection{Periodogram and Power Spectral Density}
The power spectral density (PSD) $S_{z,z}\left(\omega\right)$ of a stationary
stochastic process $z\left(t\right)$ is given by the
Fourier transform of its auto-correlation function
$C_{z,z}\left(\tau\right)=\langle z\left(t\right)z\left(t+\tau\right)\rangle$
(Wiener--Khinchin theorem). Given only a sample of the trajectory (in discrete time),
the PSD is typically estimated from the periodogram \cite{Priestley1982, Madsen2007},
which is given by the absolute square of the (discrete) Fourier transform of the time series
\begin{equation} \label{eq: periodogram}
\widehat{S}_{z,z}\left(\omega\right)=\left|\mathcal{F}\left[z\left(t\right)\right]\left(\omega\right)\right|^{2}.
\end{equation}
The PSD then follows as the expectation value of the periodogram (ensemble mean) 
\begin{equation}  \label{eq: PSD is mean of periodogram}
S_{z,z}\left(\omega\right) = \langle \widehat{S}_{z,z}\left(\omega\right)\rangle.
\end{equation}
Our goal is to estimate the FN--PSD of the free-running laser, which will be
denoted by $S_{x,x}\left(\omega\right)$ in the following. As the frequency follows
from the optical phase by differentiation with respect to time
$\omega\left(t\right)=\dot{\phi}\left(t\right)$, their PSDs are connected by
\begin{equation}
    \label{eq: phase to frequency PSD}
    S_{x,x}\left(\omega\right) = \omega^2 S_{\phi,\phi}\left(\omega\right).
\end{equation}
Note that the FN--PSD $S_{x,x}\left(\omega\right)$ is not directly observed
in the experiment, but must be reconstructed from the phase jitter
$\phi\left(t\right)-\phi\left(t-\tau_d\right)$, that can be deduced
from the slow beat note of the intensity time series corresponding
to Eq.~\eqref{eq:measuredE}.

\begin{figure}[t]
    \centering
    \includegraphics[scale=1.0]{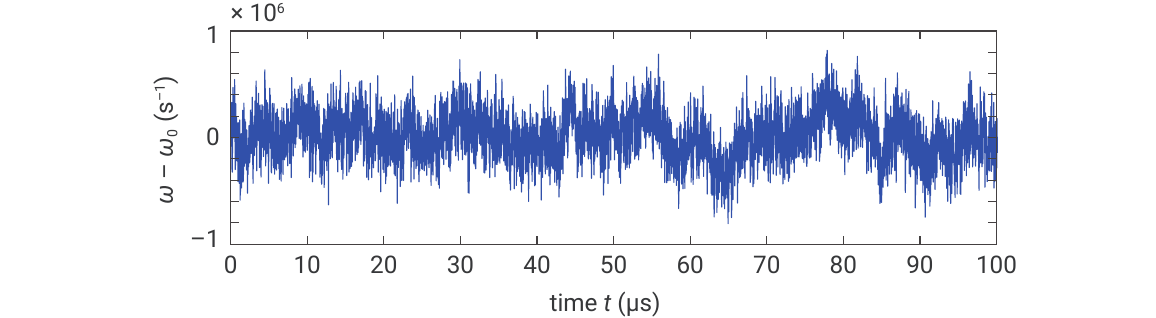}
    \caption{Simulated time series of frequency fluctuations $\omega\left(t\right)=\dot{\phi}\left(t\right)$ around the nominal continuous wave frequency $\omega_0$. The instantaneous laser frequency exhibits characteristic drifts, which are commonly observed in semiconductor lasers. The time series has been simulated using the SDE model \eqref{eq: Ito SDEs}, where frequency drifts are induced by the $1/f$--type colored noise contributions. The trajectory in the plot shows a moving average (over 50\,ns) to improve the visibility of the effect.
    }
    \label{fig:freq_drift}
\end{figure}

\begin{figure}[t]
    \centering
    \includegraphics[scale=1.0]{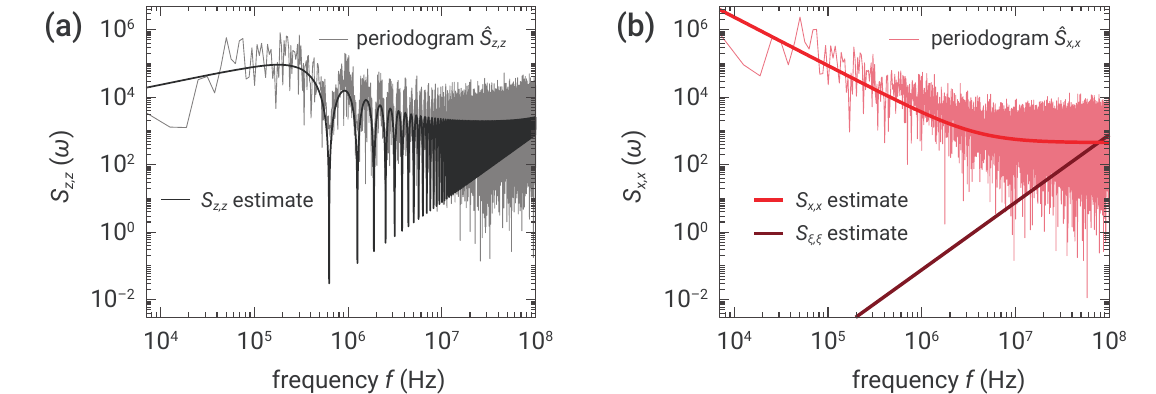}
    \caption{\textbf{(a)}~Periodogram $\widehat{S}_{z,z}\left(\omega\right)$ of an
    observed time series $z\left(t\right)$ along with the power spectral density
    $S_{z,z}\left(\omega\right)$ estimated using the MCMC method. The signal features
    sharp dropouts at the roots of the transfer function
    \eqref{eq: fourier domain transfer function H} at frequencies
    $f_n = n/\tau_d$, $n\in\mathbb{Z}$.
    \textbf{(b)}~Estimated FN--PSD $S_{x,x}\left(\omega\right)$ and
    periodogram $\widehat{S}_{x,x}\left(\omega\right)$ of the hidden time
    series $x\left(t\right)$. The detector noise PSD $S_{\xi,\xi}\left(\omega\right)$
    has a quadratic frequency dependency. Both PSDs are obtained from statistical
    inference on the observed periodogram $\widehat{S}_{z,z}\left(\omega\right)$.
    }
    \label{fig:PSDs}
\end{figure}

In the following, we describe the measured signal of the DSH experiment
(in terms of frequency fluctuations) by 
\begin{equation}
z\left(t\right)=\left(h*x\right)\left(t\right)+\xi\left(t\right),\label{eq: measurement z}
\end{equation}
where $z\left(t\right)$ is the observed time series, the convolution kernel 
$h\left(t\right)=\delta\left(t\right)-\delta\left(t-\tau_{d}\right)$ is the
transfer function of the interferometer, $x\left(t\right)$ is the hidden time series
of the instantaneous frequency fluctuations of the laser
(\emph{i.e.}, $x\left(t\right)\;\hat{=}\;\omega\left(t\right)-\omega_{0}=\dot{\phi}\left(t\right)$)
and $\xi\left(t\right)$ is (colored) additive measurement noise
(not correlated with the hidden signal).
A sample time series of frequency fluctuations $x\left(t\right)$, which exhibits
characteristic frequency drifts as commonly observed in semiconductor lasers,
is shown in Fig.~\ref{fig:freq_drift}. Our goal is to characterize the statistical
properties of the fluctuating time series $x\left(t\right)$.
Fourier transform of Eq.~\eqref{eq: measurement z} yields a relation
between the PSDs of the observed and the hidden signal
\begin{equation}
S_{z,z}\left(\omega\right) = \left|H\left(\omega\right)\right|^{2} S_{x,x}\left(\omega\right) + S_{\xi,\xi}\left(\omega\right),\label{eq: Szz equation}
\end{equation}
where the Fourier transformed transfer function reads
\begin{equation} 
    H\left(\omega\right)=1-\exp{\left(i\omega\tau_{d}\right)}. \label{eq: fourier domain transfer function H}
\end{equation}
The PSDs of the hidden signal and the detector
noise are assumed to obey the following functional forms \cite{Kantner2023}
\begin{align}
S_{x,x}\left(\omega\right) & =\frac{C}{\left|\omega\right|^{\nu}}+S_{\infty}, & S_{\xi,\xi}\left(\omega\right) & =\sigma^{2}\omega^{2}.\label{eq: Sxx and Sxixi}
\end{align}
The model equation for $S_{x,x}\left(\omega\right)$ is a phenomenological relation \cite{Kikuchi1989, Mercer1991} that includes both $1/f^\nu$--type noise (described by $C$ and $\nu$) and a white noise plateau, where $S_{\infty}$ quantifies the intrinsic linewidth. 
The detector noise PSD $S_{\xi,\xi}\left(\omega\right)$ follows from the assumption of spectrally white phase fluctuation measurement noise, 
which must be multiplied by $\omega^2$ to arrive at the corresponding  measurement noise for the frequency fluctuations, cf.\,Eq.~\eqref{eq: phase to frequency PSD}.
Figure~\ref{fig:PSDs} shows a double-logarithmic plot the PSDs considered here.

\section{Parameter Inference}
For estimation of the parameters characterizing the FN--PSD of the laser,
we perform a Bayesian regression on the PSD $S_{z,z}\left(\omega\right)$ of the
detected signal using the transfer function \eqref{eq: fourier domain transfer function H}
and the model relations \eqref{eq: Sxx and Sxixi}, which have a highly nonlinear
frequency dependency.
In the regression procedure, we regard the periodogram $\widehat{S}_{z,z}(\omega)$
(available at a set of discrete frequencies) as observed data.
In order to conduct a maximum likelihood estimation using frequency domain data, knowledge about the expected
statistical distribution of the periodogram $\widehat{S}_{z,z}(\omega)$ is required.

The frequency fluctuations $x\left(t\right)$ and the measurement noise $\xi\left(t\right)$
are assumed to normally distributed, which is in excellent agreement with
experimental data. The detected time series $z\left(t\right)$ observed at
discrete instances of time is thus a multi-variate Gaussian characterized by its
covariance matrix. By random variable transformation \cite{Priestley1982} we then
find the periodogram of the measured time series to be exponentially distributed
\begin{equation} \label{eq: exponential law for periodogram}
    \widehat{S}_{z,z}\left(\omega\right) \sim \mathrm{Exp}\left(\lambda\left(\omega, \boldsymbol{\theta}\right)\right),
\end{equation}
where the probability distribution function (PDF)
\begin{equation}
p\left(\widehat{S}_{z,z}\left(\omega\right),\lambda\left(\omega,\boldsymbol{\theta}\right)\right)=\lambda\left(\omega,\boldsymbol{\theta}\right)\mathrm{e}^{-\lambda\left(\omega,\boldsymbol{\theta}\right)\,\widehat{S}_{z,z}\left(\omega\right)},    \label{eq:pdf S_zz}
\end{equation}
is characterized by a parameter $\lambda$ that depends on frequency $\omega$
and the unknown parameters $\boldsymbol{\theta}=\left(C,\nu,S_{\infty},\sigma\right)^{T}$.
We identify the parameter function $\lambda\left(\omega, \boldsymbol{\theta}\right)$ 
with the inverse expectation value of the periodogram data given by Eq.~\eqref{eq: Szz equation} such that
\begin{equation}
    \frac{1}{\lambda\left(\omega,\boldsymbol{\theta}\right)}=S_{z,z}\left(\omega,\boldsymbol{\theta}\right)=\left|H\left(\omega\right)\right|^{2}S_{x,x}\left(\omega,\boldsymbol{\theta}\right)+S_{\xi,\xi}\left(\omega,\boldsymbol{\theta}\right).
\end{equation}
The likelihood of observing a certain realization of the periodogram $\widehat{S}_{z,z}\left(\omega_k\right)$ (at a discrete set of frequencies $\omega_k$) given a set of parameters $\boldsymbol{\theta}$ is then given by the likelihood function
\begin{equation}\label{eq:likelihood-func}
L\left(\boldsymbol{\theta}\right) = \prod_{k}
p\left(\widehat{S}_{z,z}\left(\omega_k\right),\lambda\left(\omega_k,\boldsymbol{\theta}\right)\right), 
\end{equation}
where the function $p$ is given by Eq.~\eqref{eq:pdf S_zz}.
Note that the underlying joint probability distribution factorizes here completely, as the periodogram data $\widehat{S}_{z,z}\left(\omega\right)$ at different frequencies are statistically independent.

Bayesian regression on the parameters $\boldsymbol{\theta}$ is now performed by maximizing the likelihood function \eqref{eq:likelihood-func} using a Markov chain Monte Carlo (MCMC) approach \cite{Naesseth2019}. We employ the Metropolis--Hastings algorithm to sample a Markov chain of parameter sets $\boldsymbol{\theta}^{\left(j-1\right)}\to\boldsymbol{\theta}^{\left(j\right)}\to\ldots$ that is distributed according to $L\left(\boldsymbol{\theta}\right)$.
The maximum of this distribution is then located at the parameter set $\boldsymbol{\theta}^\ast$ which is most likely to underlie the observed data.

We would like to stress that in the present scenario the regression on spectral data
(\emph{i.e.}, on the periodogram $\widehat{S}_{z,z}\left(\omega\right)$) as
described above is advantageous compared to the direct estimation on time domain data.
The reason for this is that the construction of the likelihood function for the
observed time series is either conceptually challenging due to the long
interferometer delay or would require a high-dimensional Markovian embedding.
Furthermore, the system exhibits long-range time correlations due to the
presence of $1/f$--type noise, which would require very long time series to
be considered in the regression. These issues render the estimation procedure
in the time domain computationally expensive, but can be easily overcome by
transforming the problem into the frequency domain.

\section{Results}

We demonstrate the estimation method described above for simulated data using
the stochastic laser rate equations given in
Appendix~\ref{sec: stochastic rate equations}. The DSH measurement is simulated
as described in Ref.~\cite{Kantner2023} with simulation code that is available online \cite{Kantner2023b}.

\begin{figure}[t]
    \centering
    \includegraphics[scale=1.0]{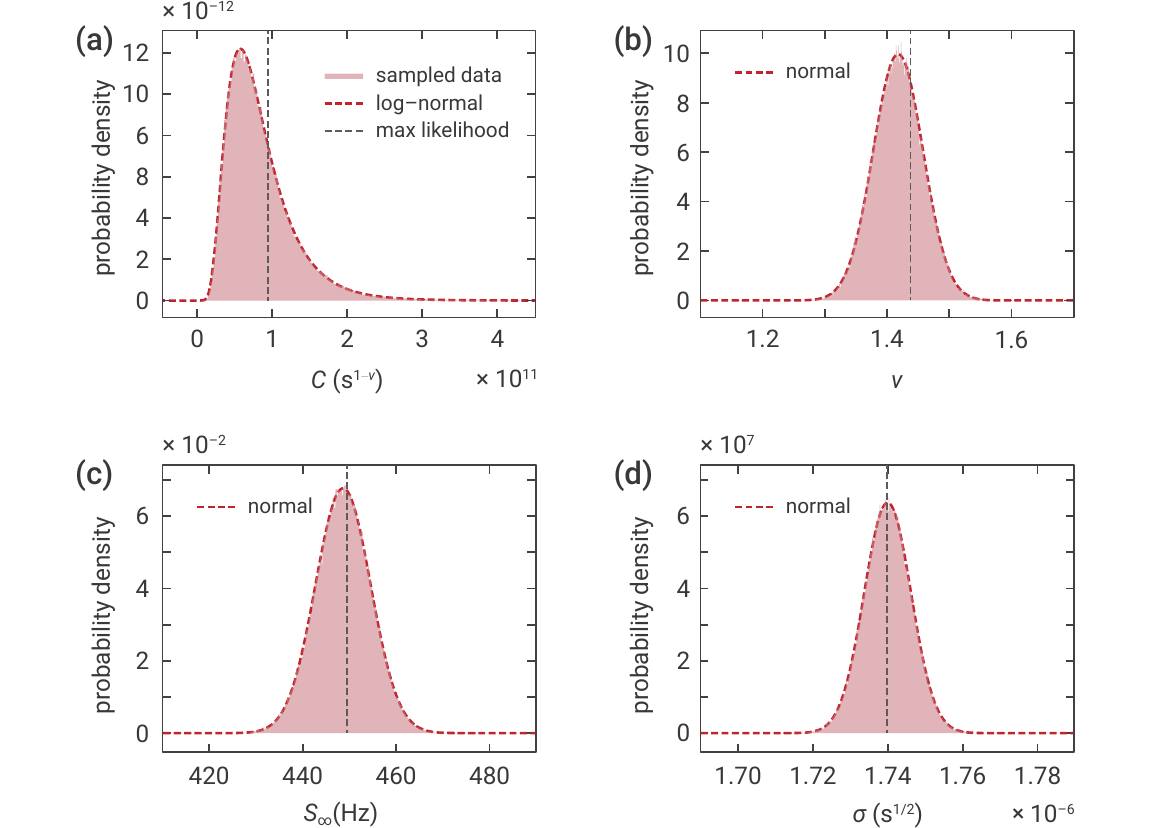}
    \caption{Histograms of estimated parameters characterizing the FN--PSD obtained using the MCMC method. The sampled distributions are shown along with normally (Gaussian) and log-normally distributed probability density distributions. 
    The dashed black line indicates the position of the maximum likelihood estimate. The estimation results are summarized in Tab.~\ref{tbl: results}.
    }
    \label{fig:histograms}
\end{figure}

Figure~\ref{fig:histograms} shows histograms of the sampled Markov chains $\boldsymbol{\theta}^{\left(j\right)}$,
which are estimators of the marginal distributions of
the joint PDF of the parameters that is proportional to $L\left(\boldsymbol{\theta}\right)$.
A notable feature of the MCMC method is that it provides not only an estimate of
the most probable set of parameters but rather their entire distribution,
allowing to asses the uncertainty of the estimates, see Tab.~\ref{tbl: results}. 
In addition to the histogram, the plot also indicates the position of the maximum
of $L\left(\boldsymbol{\theta}\right)$ as a dashed line. The corresponding parameter
set $\boldsymbol{\theta}^*$ is denoted as the maximum likelihood estimator (MLE), \emph{i.e.}, the maximum of
the joint PDF. It should be noted that these values partially differ significantly
from the means of the marginal distributions. The reason for this is that the different
parameters are strongly correlated, and the marginal distribution of $C$ has a maximum
far from its mean. The mutual correlation of all parameter estimates is quantified
by the matrix of Pearson correlation coefficients
\begin{equation} \label{eq: corr}    
    \rho_{i,j}=\frac{\mathrm{Cov}\left(\theta_{i},\theta_{j}\right)}{\sqrt{\mathrm{Var}\left(\theta_{i}\right)\,\mathrm{Var}\left(\theta_{j}\right)}},
\end{equation} 
which is obtained as
\begin{equation*}
\renewcommand{\kbldelim}{(}
\renewcommand{\kbrdelim}{)}
\rho =\quad  \kbordermatrix{
   &C&\nu&S_\infty&\sigma \\
 C&1 & 0.93 & 0.41 & -0.14\\
 \nu & & 1 & 0.47 & -0.16\\
 S_\infty & &  & 1 & -0.46\\
 \sigma & &  &  & 1
 }.
\end{equation*}
Indeed, one observes in Fig.~\ref{fig:histograms}\,(b)--(d) that the MLE and the means
of the marginal distributions differ more significantly, the stronger the respective
parameter estimate is correlated with $C$. Moreover, we observe that the estimates
of all parameters entering the signal PSD $S_{x,x}\left(\omega\right)$
are negatively correlated with the estimate of the detector noise parameter $\sigma$,
which is expected from Eq.~\eqref{eq: Szz equation}.

The MCMC routine was run with normally distributed proposal functions for the
parameters $\nu$, $S_\infty$ and $\sigma$ and with a log-normal proposal
distribution for $C$. The latter automatically enforces $C>0$ and provides an
efficient sampling as it matches the shape of the target distribution.
With the variances of the proposal distribution tuned appropriately, we achieved
an acceptance rate of the proposed parameter samples of around $30\%$.

\begin{table}[t]
\centering
\begin{tblr}{c c c}
\toprule
\textbf{parameter (unit)}         & \textbf{mean} $\mathbf{\pm}$ \textbf{std}   & \textbf{MLE} \\
\midrule
$C\,(\mathrm{s}^{1-\nu})$  & $\left(8.6\pm 4.8\right) \cdot 10^{10}$& $9.6\cdot10^{10}$\\
$\nu$                     & $1.42\pm 0.04$  &  $1.44$\\
$S_\infty\,(\mathrm{Hz})$ & $448.8\pm 5.9$ & $449.6$\\
$\sigma\,(\mathrm{s}^{1/2})$ & $1.74\cdot 10^{-6}\pm 6.4\cdot 10^{-9}$ & $1.74\cdot10^{-6}$\\
\bottomrule
\end{tblr}
\caption{Mean values and standard deviations (std) of the marginal distributions and the maximum likelihood estimator (MLE) of the parameters $\boldsymbol{\theta}$ obtained using the MCMC method.}
\label{tbl: results}
\end{table}

\section{Conclusions}
The application of Bayesian inference methods to data from DSH laser linewidth
measurements allows for an accurate extraction of the parameters characterizing the FN--PSD
along with its uncertainties. Based on statistical modeling of the underlying measurement
process, the method enables a reliable characterization of the spectral coherence
of the laser even when the intrinsic linewidth plateau is obscured by detector noise.
Furthermore, by invoking explicit assumptions about the statistical distribution of
the measured data, the method extracts the information encoded in the available
data very extensively, so that no averaging over many realizations or long times is required.

\appendix

\section*{Appendix}
\setcounter{section}{0} 

\renewcommand \thesection{\Alph{section}}

\section{Stochastic Laser Rate Equations} \label{sec: stochastic rate equations}
We describe a set of Itô-type stochastic differential equations (SDEs) modeling the fluctuation dynamics of a generic single-mode semiconductor laser. In the presence of noise, the evolution of the photon number $P$, the optical phase $\phi$ and the carrier number $N$ in the active region is described by:
\begin{subequations}\label{eq: Ito SDEs}
\begin{align}
\mathrm{d}P&=
\left(
-\gamma\left(P-P_{\mathrm{th}}\right)
+\Gamma v_{g}g\left(P,N\right)P
+\Gamma v_{g}g_{\mathrm{sp}}\left(P,N\right)
+\sigma_{P}\left(P\right)\mathcal{F}_{P}\left(t\right)
\right)\,\mathrm{d}t\label{eq:SDEP}\\
&\hphantom{=}
+\sqrt{\gamma\left(1+P_{\mathrm{th}}\right)P}\,\mathrm{d}W_{\mathrm{out}}^{P}
+\sqrt{\gamma P_{\mathrm{th}}\left(1+P\right)}\,\mathrm{d}W_{\mathrm{in}}^{P}
+\sqrt{\Gamma v_{g}g_{\mathrm{sp}}\left(P,N\right)}\,\mathrm{d}W_{\mathrm{sp}}^{P}
\nonumber\\
&\hphantom{=}
+\sqrt{\Gamma v_{g}g_{\mathrm{sp}}\left(P,N\right)P}\,\mathrm{d}W_{\mathrm{st-em}}^{P}
+\sqrt{\Gamma v_{g}g_{\mathrm{abs}}\left(P,N\right)P}\,\mathrm{d}W_{\mathrm{st-abs}}^{P},
\nonumber\\
\mathrm{d}\phi&=
\left(
\Omega_{0}
+\frac{\alpha_{H}}{2}\Gamma v_{g}g\left(P,N\right)
+\frac{\sigma_{P}\left(P\right)}{2P}\mathcal{F}_{\phi}\left(t\right)
\right)\,\mathrm{d}t\label{eq:SDEphi}\\
&\hphantom{=}
+\frac{1}{2P}\bigg(
\sqrt{\gamma\left(1+P_{\mathrm{th}}\right)P}\,\mathrm{d}W_{\mathrm{out}}^{\phi}
+\sqrt{\gamma P_{\mathrm{th}}\left(1+P\right)}\,\mathrm{d}W_{\mathrm{in}}^{\phi}
+\sqrt{\Gamma v_{g}g_{\mathrm{sp}}\left(P,N\right)}\,\mathrm{d}W_{\mathrm{sp}}^{\phi}
\nonumber\\
&\hphantom{=+\frac{1}{2P}\bigg(}
+\sqrt{\Gamma v_{g}g_{\mathrm{sp}}\left(P,N\right)P}\,\mathrm{d}W_{\mathrm{st-em}}^{\phi}
+\sqrt{\Gamma v_{g}g_{\mathrm{abs}}\left(P,N\right)P}\,\mathrm{d}W_{\mathrm{st-abs}}^{\phi}
\bigg),\nonumber\\
\mathrm{d}N&=\left(\frac{\eta I}{q}-R\left(N\right)-\Gamma v_{g}g\left(P,N\right)P-\Gamma v_{g}g_{\mathrm{sp}}\left(P,N\right)+\sigma_{N}\left(N\right)\mathcal{F}_{N}\left(t\right)\right)\,\mathrm{d}t\label{eq:SDEN}\\&\hphantom{=}+\sqrt{\frac{\eta I}{q}}\,\mathrm{d}W_{I}+\sqrt{R\left(N\right)}\,\mathrm{d}W_{R}-\sqrt{\Gamma v_{g}g_{\mathrm{sp}}\left(P,N\right)P}\,\mathrm{d}W_{\mathrm{st-em}}^{P} \nonumber\\
&\hphantom{=} -\sqrt{\Gamma v_{g}g_{\mathrm{abs}}\left(P,N\right)P}\,\mathrm{d}W_{\mathrm{st-abs}}^{P}-\sqrt{\Gamma v_{g}g_{\mathrm{sp}}\left(P,N\right)}\,\mathrm{d}W_{\mathrm{sp}}^{P}.\nonumber
\end{align}
\end{subequations}
Here, $\gamma$ is the optical loss rate, $P_{\mathrm{th}}$ is the thermal photon
number (Bose--Einstein factor), $\Gamma$ is the optical confinement factor,
$v_g$ is the group velocity, $\Omega_0$ is a detuning from the nominal CW
frequency, $\alpha_H$ is the linewidth enhancement factor (Henry factor),
$\eta$ is the injection efficiency, $I$ is the pump current and
$q$ is the elementary charge. The net-gain is modeled as
\begin{equation*} \label{eq:net-gain}
g\left(P,N\right)=\frac{g_{0}}{1+\varepsilon P}\log\left(\frac{N}{N_{\mathrm{tr}}}\right),
\end{equation*}
where $g_0$ is the gain coefficient, $\varepsilon$ is the inverse saturation
photon number (modeling gain compression) and $N_{\mathrm{tr}}$ is the 
carrier number at transparency. Following \cite{Wenzel2021}, the rate of spontaneous
emission into the lasing mode is modeled as
\begin{equation*} \label{eq:spont-em}
g_{\mathrm{sp}}\left(P,N\right)=\frac{1}{2}\frac{g_{0}}{1+\varepsilon P}\log\left(1+\left(\frac{N}{N_{\mathrm{tr}}}\right)^{2}\right),
\end{equation*}
which does not require any additional parameters and shows the correct asymptotics at low
and high carrier numbers. The rate of stimulated absorption entering the noise amplitudes
follows as $g_{\mathrm{abs}}\left(P,N\right)=g_{\mathrm{sp}}\left(P,N\right)-g\left(P,N\right)$.
Finally, non-radiative recombination and spontaneous emission into waste modes is
described by
\begin{equation*} \label{eq:reco}
R\left(N\right)=AN+\frac{B}{V}N^{2}+\frac{C}{V^{2}}N^{3}.
\end{equation*}
We refer to Ref.~\cite{Kantner2023} for a list of parameter values used in the simulation.

The model equations \eqref{eq: Ito SDEs} include white and colored noise contributions.
Here, $\mathrm{d}W\sim\mathrm{Normal}\left(0,\mathrm{d}t\right)$
denotes the increment of the standard Wiener processes  
modeling Gaussian white noise. Wiener processes with different sub- and superscripts are statistically independent.
Colored noise sources $\mathcal{F}_{P,\phi,N}\left(t\right)$ are constructed as superpositions of Ornstein--Uhlenbeck processes, where the parameters are calibrated to result in power spectral densities showing a $1/f^\nu$--type frequency dependency.
See Ref.~\cite{Kantner2023} for details.

\begin{acknowledgement}
This work was funded by the German Research Foundation (DFG) under Germany's Excellence Strategy – EXC2046: MATH+ (project AA2-13).
\end{acknowledgement}

\footnotesize
\bibliographystyle{spmpsci_unsort}

\begin{thebibliography}{10}
\providecommand{\url}[1]{{#1}}
\providecommand{\urlprefix}{URL }
\expandafter\ifx\csname urlstyle\endcsname\relax
  \providecommand{\doi}[1]{DOI~\discretionary{}{}{}#1}\else
  \providecommand{\doi}{DOI~\discretionary{}{}{}\begingroup
  \urlstyle{rm}\Url}\fi

\bibitem{Abbott2009}
Abbott, B.P., Abbott, R., Adhikari, R., \emph{et al.}: {LIGO}: the laser
  interferometer gravitational-wave observatory.
\newblock Rep. Prog. Phys. \textbf{72}(7), 076901 (2009).
\newblock \doi{10.1088/0034-4885/72/7/076901}

\bibitem{Ludlow2015}
Ludlow, A.D., Boyd, M.M., Ye, J., Peik, E., Schmidt, P.O.: Optical atomic
  clocks.
\newblock Rev. Modern Phys. \textbf{87}(2), 637--701 (2015).
\newblock \doi{10.1103/RevModPhys.87.637}

\bibitem{Kikuchi2016}
Kikuchi, K.: Fundamentals of coherent optical fiber communications.
\newblock J. Lightwave Technol. \textbf{34}(1), 157--179 (2016).
\newblock \doi{10.1109/JLT.2015.2463719}

\bibitem{Akerman2015}
Akerman, N., Navon, N., Kotler, S., Glickman, Y., Ozeri, R.: Universal gate-set
  for trapped-ion qubits using a narrow linewidth diode laser.
\newblock New. J. Phys. \textbf{17}(11), 113060 (2015).
\newblock \doi{10.1088/1367-2630/17/11/113060}

\bibitem{Kikuchi1989}
Kikuchi, K.: Effect of 1/f-type {FM} noise on semiconductor-laser linewidth
  residual in high-power limit.
\newblock IEEE J. Quant. Electron. \textbf{25}(4), 684--688 (1989).
\newblock \doi{10.1109/3.17331}

\bibitem{Mercer1991}
Mercer, L.B.: 1/f frequency noise effects on self-heterodyne linewidth
  measurements.
\newblock J. Lightwave Technol. \textbf{9}(4), 485--493 (1991).
\newblock \doi{10.1109/50.76663}

\bibitem{Stephan2005}
St{\'{e}}phan, G.M., Tam, T.T., Blin, S., Besnard, P., T{\^{e}}tu, M.: Laser
  line shape and spectral density of frequency noise.
\newblock Phys. Rev. A \textbf{71}(4), 043809 (2005).
\newblock \doi{10.1103/PhysRevA.71.043809}

\bibitem{Henry1986}
Henry, C.: Phase noise in semiconductor lasers.
\newblock J. Lightwave Technol. \textbf{4}(3), 298--311 (1986).
\newblock \doi{10.1109/JLT.1986.1074721}

\bibitem{Wenzel2021}
Wenzel, H., Kantner, M., Radziunas, M., Bandelow, U.: Semiconductor laser
  linewidth theory revisited.
\newblock Appl. Sci. \textbf{11}(13), 6004 (2021).
\newblock \doi{10.3390/app11136004}

\bibitem{Okoshi1980}
Okoshi, T., Kikuchi, K., Nakayama, A.: Novel method for high resolution
  measurement of laser output spectrum.
\newblock Electron. Lett. \textbf{16}(16), 630 (1980).
\newblock \doi{10.1049/el:19800437}

\bibitem{Schiemangk2014}
Schiemangk, M., Spie{\ss}berger, S., Wicht, A., Erbert, G., Tr\"{a}nkle, G.,
  Peters, A.: Accurate frequency noise measurement of free-running lasers.
\newblock Appl. Optics \textbf{53}(30), 7138 (2014).
\newblock \doi{10.1364/AO.53.007138}

\bibitem{Kantner2023}
Kantner, M., Mertensk\"{o}tter, L.: Accurate evaluation of self-heterodyne
  laser linewidth measurements using {Wiener} filters.
\newblock Opt. Express \textbf{31}(10), 15994--16009 (2023).
\newblock \doi{10.1364/OE.485866}

\bibitem{Priestley1982}
Priestley, M.B.: Spectral analysis and time series.
\newblock Academic Press, London (1982)

\bibitem{Madsen2007}
Madsen, H.: Time Series Analysis.
\newblock Chapman and Hall, New York (2007).
\newblock \doi{10.1201/9781420059687}

\bibitem{Naesseth2019}
Naesseth, C.A., Lindsten, F., Sch\"{o}n, T.B.: Elements of sequential {Monte}
  {Carlo}.
\newblock Foundations and Trends in Machine Learning \textbf{12}(3), 307--392
  (2019).
\newblock \doi{10.1561/2200000074}

\bibitem{Kantner2023b}
Kantner, M., Mertensk\"{o}tter, L.: {Laser Noise} ({GitHub} repository).
\newblock URL: https://github.com/kantner/LaserNoise (2023)

\end{thebibliography}

\end{document}